\documentclass[aps,prl,twocolumn,showpacs,10pt]{revtex4-1}
\usepackage{graphicx}
\usepackage{amssymb}
\usepackage{amsmath}
\usepackage{comment}
\usepackage{footmisc}
\usepackage{color}
\pdfoutput=1

\begin{document}

\title{Fractional Charge and Spin States in Topological Insulator Constrictions}

\author{Jelena Klinovaja}
\affiliation{Department of Physics, University of Basel,
            Klingelbergstrasse 82, CH-4056 Basel, Switzerland}
\author{Daniel Loss}
\affiliation{Department of Physics, University of Basel,
            Klingelbergstrasse 82, CH-4056 Basel, Switzerland}
\begin{abstract}
We investigate theoretically properties of two-dimensional topological insulator constrictions both in the integer and fractional regimes. In the presence of a perpedicular magnetic field, the constriction functions as a spin filter with near-perfect efficiency and can be switched by electric fields only. Domain walls between different topological phases can be created in the constriction as an interface between tunneling, magnetic fields, charge density wave, or electron-electron interactions dominated regions. These domain walls host non-Abelian bound states with fractional charge and spin and result in degenerate ground states with parafermions. If a proximity gap is induced bound states give rise to an exotic Josephson current with $8\pi$-peridiodicity.
\end{abstract}

\date{\today}
\pacs{71.10.Pm; 05.30.Pr; 72.25.-b
}

\maketitle

\begin{figure}[!b]
\includegraphics[width=\linewidth]{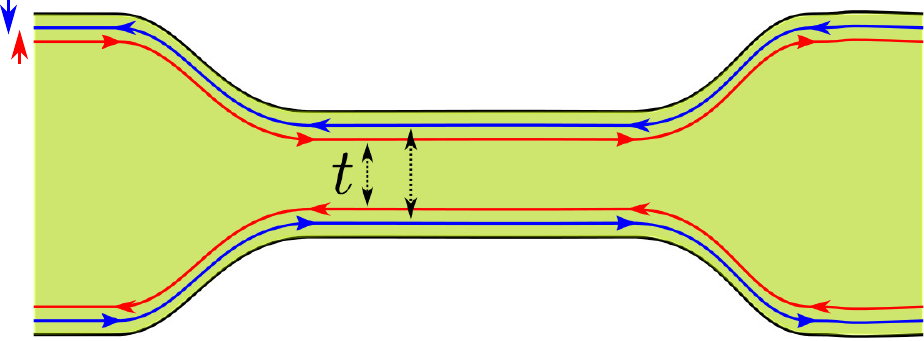}
\caption{A sketch of a constriction formed in a topological insulator. Spin up (down) edge states are shown in red (blue). The constriction could be either doped with magnetic impurities or subjected to a magnetic field.
The pairs of helical edge states are coupled by tunneling which results in the opening of an energy gap at zero momentum.
A magnetic field applied in the plane of the TI couples edge states with opposite spins and also opens a gap in the spectrum.}
\label{model}
\end{figure}

\textit{Introduction.}  
The field of topological properties in condensed matter systems has been rapidly growing over the past decade. In particular, the topics of  topological insulators (TIs) \cite{Hasan_review,Volkov_TI1,Volkov_TI2,Fu_Kane,Zhang_TI,exp_1,Konig_2 , Konig, Roth_TI,Nowack_TI,Amir_TI,Patric_TI,Daniel_Yaroslav_PRL_TI_CAS,Patric_TI_2}  and  exotic bound states with non-Abelian statistics have attracted a lot of attention theoretically and experimentally~\cite{Read_2000,fu,Nagaosa_2009,Sato,
lutchyn_majorana_wire_2010,Rotating_field,RKKY_Basel,
RKKY_Simon,RKKY_Franz,Klinovaja_CNT,bilayer_MF_2012,MF_nanoribbon,MF_MOS,MF_ee_Suhas,mourik_signatures_2012,deng_observation_2012,
das_evidence_2012,Rokhinson,Goldhaber,marcus_MF,Ali,PF_Linder,PF_Clarke,PF_Cheng,
Ady_FMF,PF_Mong,vaezi_2,PFs_Loss,PFs_Loss_2,PFs_TI,oreg_majorana_wire_2010,barkeshli_2}.
Of special interest are also quantum effects arising from  geometric confinement  such as 
topological insulator constrictions (TICs)~\cite{TI_constriction_Niu,TI_constriction_Kane,TI_constriction_2,Richter_constriction,Chang_TI,TI_constriction}, 
where the edge modes get coupled by tunneling and a gap is opened in the energy spectrum, see Fig.~\ref{model}. In general, such a coupling is not desirable since  typically it leads to
a suppression of topological properties~\cite{Hasan_review,fu}.  However, we will find that, quite surprisingly,  in the presence of additional mode-mixing perturbations such as magnetic fields, superconductivity, and interaction effects, the tunneling does not necessarily destroy all such properties. Instead, different topological phases  can emerge  that give rise to exotic phenomena such as fractional fermions, parafermions, and exotic superconductivity 
where Cooper pairs themselves get paired.

First, we consider constrictions with extended edge modes and show that in the presence of  tunneling the TIC  can be tuned between insulating, propagating, and spin filtering regimes by electric fields only, making such TICs attractive candidates for spintronics applications.

Second, we focus on localized modes. Here, we identify competing mechanisms that generate gaps in the spectrum, arising from  magnetic fields, tunneling between edges, periodic modulations of the chemical potential, proximity effects, and electron-electron interactions. We show that there are   zero-energy bound states at the interfaces between two phases controlled by competing gap mechanisms. These states are  fractional fermions of the Jackiw-Rebbi type. If fractional TIs with fractional charge $e/m$ are considered, the ground state is $m$-fold degenerate and the resulting bound states are $\mathbb{Z}_m$-parafermions. The superconductivity that could be induced by proximity effect at such  constrictions corresponds to the coherent tunneling of two Cooper pairs and results in an unusual $8\pi$-periodic Josephson current.

\textit{TIC Model. }
We consider a constriction created in a two-dimensional TI, see Fig. 
\ref{model}. Upper and lower edges of the TIC hosting helical states of opposite helicities are brought close to each other and, as a result, couple via tunneling. The edges of the TIC are labeled by the index  $\tau$, where $\tau=1$ ($\tau=-1$) corresponds to the upper (lower) edge. The helical edge states of the TIC have a linear energy dispersion. The corresponding kinetic part of the Hamiltonian is given by $H_{kin}=-i\hbar \upsilon_F \sum_\tau ( R_\tau^\dagger \partial_x R_\tau-L_\tau^\dagger \partial_x L_\tau)$,
where $\upsilon_F$ is the Fermi velocity. The operator $ R_\tau (x)$ [$ L_\tau (x)$] is the annihilation operator acting on the right-propagating (left-propagating) electron located at point $x$ of the TIC edge  $\tau$. 
We note here that the two pairs of helical edge states possess opposite helicities, {\it i.e.,} right-propagating (left-propagating) electrons at the upper (lower) edge are spin-up electrons and left-propagating (right-propagating) electrons at the upper (lower) edge are spin-down electrons, see Fig.~ \ref{model}. 

We point out that the same setup could be assembled by bringing close to each other two TI samples \cite{PFs_TI} or in the framework of strip of stripes models \cite{yaroslav}. The latter is especially important for the fractional regime \cite{Lebed,Kane_PRL,Stripes_PRL,Kane_PRB,Stripes_arxiv,Stripes_nuclear,Neupert,
tobias_1,Oreg,yaroslav,gefen} as it allows one to design fractional TIs for an array of coupled one-dimensional channels with spin-orbit interaction \cite{yaroslav}.

The tunneling in the TIC is assumed to be spin conserving and described by
$H_{tun}= \sum_\sigma \Psi^\dagger_{1\sigma} \Psi_{\bar 1\sigma} + H.c.$,
where $t$ is the  tunneling matrix element between two edges of the TI and the operator $\Psi_{\tau \sigma}$ is the  electron annihilation operator at position $x$ of the edge state $\tau$. In what follows, we use the fact that the fast oscillating part of the wavefunction is given  by $e^{\pm k_F x}$, where $k_F$ is the Fermi wavevector set by the chemical potential $\mu=\hbar \upsilon_F k_F$. Keeping only slowly varying terms in the Hamiltonian \cite{Rotating_field,Braunecker,Klinovaja2012,MF_SOI}, we arrive at
\begin{align}
H_{tun}= t  (R_1^\dagger L_{\bar 1} + L_1^\dagger R_{\bar 1} + H.c.).
\end{align}

The TI surface could be subjected to a magnetic field or doped with magnetic impurities producing a local effective magnetic field. The  Hamiltonian is given by
$H_{Z} =  \sum_{\sigma,\sigma'}  \Delta_{n} \Psi^\dagger_{1\sigma} ({\bf n}\cdot {\boldsymbol \sigma})_{\sigma\sigma'} \Psi_{\bar 1\sigma'}$,
where the unit vector ${\bf n}$ points along the field and $ {\boldsymbol \sigma}$ is a vector composed of Pauli matrices acting on the electron spin.
For fields along the spin quantization axis of the edge states which is chosen, say,  in the $z$-direction, the corresponding effective Zeeman term is given in terms of right and left movers as
\begin{align}
H_{z} = \Delta_{z} (R_1^\dagger R_{ 1} -L_1^\dagger L_{ 1}-R_{\bar 1}^\dagger R_{ \bar 1} + L_{\bar 1}^\dagger L_{ \bar 1}+ H.c.),
\end{align}
where $\Delta_{z} $ is the coupling constant either determined by the Zeeman energy or by the strength of exchange interaction.
If $\Delta_{z}$ is generated by a magnetic field $B_z$ with the vector potential ${\bf A }= B_z y \hat{x}$ applied perpendicular to the TI plane, 
then the wavevector $k$ gets shifted to $k -(e/\hbar c) A_x$, accounting for orbital effects of the magnetic field \cite{TI_constriction_Niu}.
If the upper (lower) edge state is at $y=d$  ($y=-d$), the shift is given by $ - \tau (e/\hbar c) d $. Interestingly, for TI edge states, the orbital and spin contributions add up to
$\Delta_{z} - e B_z d \upsilon_F / c $. However,  for typical TIC sizes one can neglect the orbital part.

The magnetic field applied perpendicular to the spin polarization axis, say, in the $x$ direction, results in the Hamiltonian
\begin{align}
H_{x} = \Delta_{x} (R_1^\dagger L_{1}+R_{\bar 1}^\dagger L_{ \bar 1} + H.c.),
\label {htau}
\end{align}
where $\Delta_{x}$ is the strength of coupling in the $x$ direction.

\begin{figure}[!b]
\includegraphics[width=\linewidth]{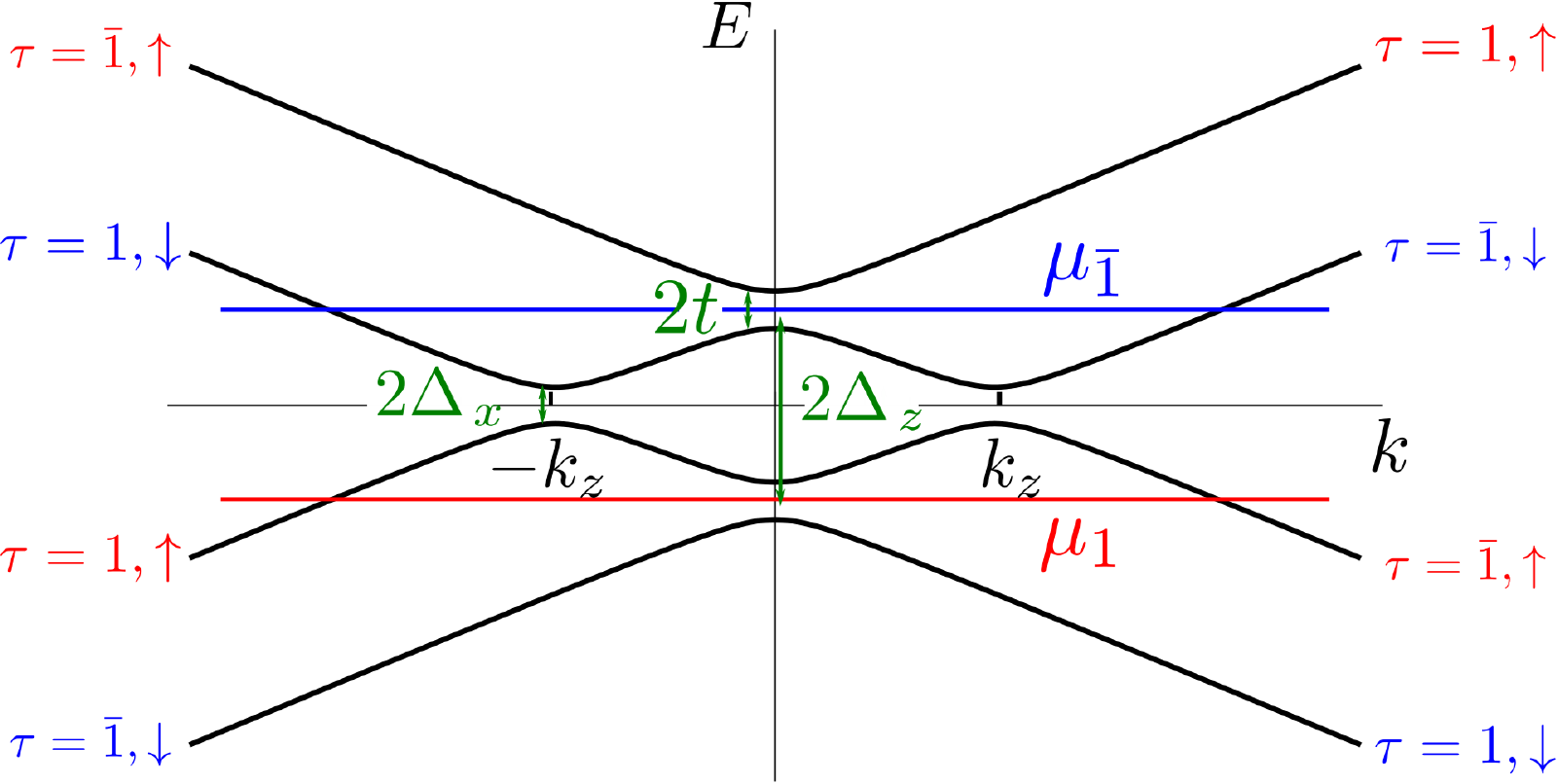}
\caption{The energy spectrum of the TIC in the presence of magnetic fields (or exchange interactions). A magnetic field in $z$-direction, the spin quantization axis of the TIC, shifts the Dirac cone corresponding to the upper (lower) edge, $\tau=1$ ($\tau=\bar 1$), to the left (right)  such that the Dirac point is located now at wavevector $-k_Z$ ($k_{Z}$). The tunneling between edges opens a gap of  size $2t$ at  $k=0$.  A magnetic field in $x$- direction, {\it i.e.}, perpendicular to the spin quantization axis, opens a gap of the size $2\Delta_{x}$ at zero energy. If the chemical potential $\mu_1$ ($\mu_{\bar 1}$) is tuned inside the gap opened by tunneling at $k=0$, only spin-polarized modes with spin up (spin down)  propagate through the TIC.}
\label{spectrum}
\end{figure}

\textit{Spin filter effect.}  In the presence of both tunneling and magnetic fields, the total Hamiltonian is given by $H=H_{kin}+H_{tun}+H_{z}+H_{x}$ and can be rewritten in the basis $(R_1, L_1, R_{\bar 1}, L_{\bar 1})$ in terms of Pauli matrices as
\begin{align}
\mathcal{H} =  \upsilon_F \hat p \rho_3 + \Delta_{x} \rho_1  +\Delta_{z} \rho_3 \tau_3 + t \rho_1 \tau_1,
\end{align}
where $ \hat p = -i\hbar \partial_x$ is the momentum operator and, for simplicity, we assume that $\Delta_{x}$, $\Delta_{z}$, and $t$ are non-negative if not specified otherwise. The Pauli operators $\rho_i$ act in right/left mover space. The energy spectrum is given by
\begin{align}
E_{\pm}^2=&(\hbar \upsilon_F k)^2 +t^2+\Delta_{z}^2+\Delta_{x}^2 \nonumber \\
&\hspace{35 pt}\pm 2 \sqrt{\Delta_{x}^2 t^2 + [(\hbar \upsilon_F k)^2+t^2]\Delta_{z}^2}.
\label{spectrum_eq}
\end{align}
We are interested in the regime $\Delta_{z}>t>\Delta_{x}$.  First, we notice that the two Dirac  cones are shifted by the perpendicular magnetic field  by $k_{z} = \Delta_{z}/\hbar \upsilon_F$ to the left (right) for the upper (lower) edge. The tunneling opens a gap at zero momentum  $k=0$ of the size  $\Delta_{k=0}=2t$, see Fig. \ref{spectrum}, while the magnetic field in the $x$ direction opens a gap at finite momentum $k=\pm k_z$ and at zero energy, given by $\Delta_{k=\pm k_z}=2 \Delta_{x}$, see Fig. \ref{spectrum}.

The described setup can be used as a spin filter controlled purely by electric gates.
For example, if $\Delta_{x}=0$, the spin projection on the $z$ axis, $s_z$, is a good quantum number and all modes are spin polarized. If the chemical potential lies in the electron (hole) part of the spectrum, $\mu \in (\Delta_{z}-t, \Delta_{z}+t)$ [$\mu \in (-\Delta_{z}-t, -\Delta_{z}+t)$], only the spin down (spin up) component can propagate through the TIC, see Fig. \ref{spectrum}. However, due to the tunneling there is leakage from the upper  to the lower edge such that the probability to stay in the upper edge is given by $\left<\tau_z\right>|_{s_z} \approx 1-(t/4\Delta_{z})^2$.
If $\mu$ is tuned close to zero, $|\mu| < \Delta_{x}$, the system is fully insulating. For other values of $\mu$ both spin components can propagate.

If $\Delta_{x}\neq 0$, the propagating modes are no longer  perfectly spin-polarized, however, deviations are small in the ratio $\Delta_{x}/\Delta_{z}$. The chemical potential should be tuned into the window of $\mu \in (\sqrt{\Delta_{z}^2+\Delta_{x}^2}-t, \sqrt{\Delta_{z}^2+\Delta_{x}^2}+t)$ [$\mu \in (-\sqrt{\Delta_{z}^2+\Delta_{x}^2}-t, -\sqrt{\Delta_{z}^2+\Delta_{x}^2}+t)$] for the spin down (up) dominated propagation, see Fig.~\ref{spectrum}.
The efficiency of the spin filter is characterized by the probability to keep an initial spin polarization and to stay at the initial edge, which is given by
\begin{align}
\left<\tau_z\right>|_{s_z} \approx 1-(t/4\Delta_{z})^2- (\Delta_{x}/2\Delta_{z})^2.
\end{align}

By changing the position of the chemical potential, {\it i.e.}, by applying {\it electric fields}, one can  tune the TIC into different spin filtering regimes. This provides a substantial  advantage over  spin filters tuned by magnetic fields which are difficult to switch fast and locally.

{\it Bound states at the tunneling-magnetic field interface.} The TIC with spectrum Eq. (\ref {spectrum_eq}) not only allows one to realize spin filtering but also to trap bound states that are localized at the interface between tunneling- and magnetic field-dominated regions.
While the energy branch $E_+$ is always gapped, the branch $E_-$ is gapless at zero momentum if $t^2=\Delta_z^2+\Delta_x^2$. At other values $E_-$ is gapped unless $\Delta_x=0$. We note that, generally, the interface separating two regions that are characterized by opposite signs of the expression $t^2-\Delta_z^2-\Delta_x^2$ hosts zero-energy bound states. As an example, we consider an interface at the left end of the TIC ($x=0$) specified by $t=0$ for $x<0$ and  by $t>\Delta_x$ with $\Delta_z=0$ for $x>0$. This interface hosts a zero-energy bound state with wavefunction of the form $\Phi(x)=(f, -i f,f^*, i f^*)^T$ with $f(x)=i \theta(x) e^{-x/\xi_t} +i \theta(-x)   e^{i k_{z} x} e^{-x/\xi_x}$,
where the localization lengths are defined as $\xi_t=\hbar \upsilon_F/(t-\Delta_x)  $ and $\xi_x=\hbar \upsilon_F/\Delta_x$. An analogous  bound state occurs also at the right end of the constriction. These zero-energy bound states are examples of fractional fermions of the Jackiw-Rebbi type~\cite{JR_model,CDW_suhas,SSH_model,FF_non_Abelian,FF_transport,FF_pump} and possess non-Abelian braiding statistics \cite{FF_non_Abelian}.

In passing we note that, alternatively, degenerate bound states even occur for $t=\Delta_z=0$, namely in the presence of a magnetic domain wall separating two domains with $\Delta_x(x)=\theta(x) \Delta_x -\theta(-x) \Delta_x$.
Such an interface hosts a zero-energy bound state  at each of the two TI edge states. The corresponding wavefunction is $\Phi(x)=(i, 1)^T e^{-|x|/\xi_B}$, where $\xi_B=\hbar \upsilon_F/\Delta_{x}$. We note that as the gap closes twice (at the upper and at the lower edge), the twofold degeneracy is not protected and states split away from zero if the tunneling is included. If the magnetization rotates not exactly by $\pi$ but by some finite angle $\chi$, the bound state moves away from zero energy, $E=-\Delta_{x} \cos(\chi/2)$ for $0<\chi <2 \pi$. The domain wall localizes the charge $e/2$ only for $\chi=\pi$, which brings us back to the fractional fermions of the Jackiw-Rebbi type~\cite{JR_model,CDW_suhas,SSH_model,FF_non_Abelian,FF_transport,FF_pump}.

{\it  Bound states at charge density wave - magnetic field interface.}
An alternative way to generate bound states is to allow for  modulations of the chemical potential $\mu_{mod}$ with the period of $2k_{Z}$,
$\mu_{mod}= 2 \delta  \mu  \cos (2k_{Z} x +\phi)$,
where $2 \delta \mu $ is the amplitude of modulations and $\phi$ is the phase at $x=0$. 
This creates a charge density wave (CDW) that opens a gap around the Fermi points. This setup works in the spin-filtering regime $\Delta_z\gg t,\Delta_x$. 
Indeed, the corresponding Hamiltonian is given by
\begin{equation}
H_{mod} = \bar { \mu} ( e^ {i\phi} R_1^\dagger L_{\bar 1}+ e^ {i\phi} L_{ 1}^\dagger R_{ \bar 1} + H.c.),
\label{hc}
\end{equation}
where the coupling amplitude is found in second order perturbation theory as $ \bar { \mu} \approx t  \delta  \mu  /\Delta_{z}$.
The energy spectrum then becomes
$E_{\pm}^2 =  (\hbar \upsilon_F k)^2 +(\Delta_{x}\pm \bar { \mu})^2$.
The bulk gap closes if $\Delta_{x}= \bar { \mu}$, indicating the topological phase transition. Thus, we can construct an interface between the magnetic field dominated region with $\bar{ \mu} =0$ ($x<0$) and the CDW dominated region with $\Delta_{x}=0$ ($x>0$). Again, such an interface hosts a zero-energy bound state with wavefunction of the form $\Phi(x)=(f, if, f^*, -i f^*)^T$ with $f(x)=i  e^{i  \phi/2} e^{i k_{orb} x} [\theta(x) e^{-x/\xi_>} +\theta(-x) e^{x/\xi_<}]$
in the basis of $(\Psi_{11},\Psi_{1\bar1},\Psi_{\bar 11},\Psi_{\bar 1\bar1})$. Here, the localization lengths are defined as  $\xi_>=\hbar \upsilon_F /\bar { \mu} $ and $\xi_<=\hbar \upsilon_F /\Delta_{x}$. 

{\it Fractional bound states at the charge density wave - magnetic field interface.}
Next, we consider helical edges of a {\it fractional} TI constriction with elementary excitations of charge $e/m$ defined in terms of chiral bosonic fields  $\phi_{rn}$ \cite{Ady_FTI}. First, we note that also in this regime the system could be operated as a spin filter for quasiparticles. Second, we focus on properties of domain walls in this system. The electron operators are then rewritten as $R_n=e^{im\phi_{1n}}$ and $L_n=e^{im\phi_{\bar1 n}}$.  To satisfy the anticommutation relations between original fermionic operators, we work with the following non-zero commutators for the bosonic fields,
\begin{align}
&[\phi_{rn}(x),\phi_{r'n'}(x')]=\frac{i\pi r}{m} \delta_{rr'}\delta_{nn'} {\rm sgn} (x-x').\label{com1}
\end{align}
All other commutators are assumed to vanish. 
To proceed, we bosonize the magnetic field Hamiltonian
$H_{x}$  [see Eq.~(\ref{htau})] as $H_{x} =2 \Delta_{x} \big(\cos [m(\phi_{11}-\phi_{\bar 1 1})]+  \cos [m(\phi_{ 1\bar1}-\phi_{\bar 1 \bar 1})]\big)$ and the CDW Hamiltonian $H_{mod}$  [see Eq.~(\ref{hc})] as $H_{mod} =2\bar \mu \big(\cos [m(\phi_{11}-\phi_{\bar 1 \bar 1}-\phi)]+\cos [m(\phi_{1 \bar 1}-\phi_{\bar  1 1}+\phi)]\big)$.
In a next step, we express the chiral fields $\phi_{rn}$  in terms of their conjugated $\phi_\rho$ and $\theta_\rho$ fields ($\rho=\pm 1$) defined as 
$\phi_{rn}=[\phi_1/m + r \theta_1/m+ n(\phi_{\bar 1} + r \theta_{\bar 1} )]/2$.
The commutation relations between the newly introduced fields are given by 
$[\phi_1 (x), \theta_1 (x')]=i\pi m\ {\rm sgn} (x-x')$ and $[\phi_{\bar1} (x), \theta_{\bar 1} (x')]=i(\pi/m) {\rm sgn} (x-x')$, while all other commutators vanish. The charge density is given by $\rho(x)= \partial_x \theta_1/\pi $ and the spin density by  $s_z(x)=  \partial_x \phi_{\bar 1}/\pi$. Here, we measure charge (spin) in units of the quasi-particle charge $e/m$ (of the electron spin $\hbar/2$). 
The non-quadratic parts of the Hamiltonian  become
\begin{align}
&H_{x} =4 \Delta_{x} \cos (\theta_1)\cos (m\theta_{\bar 1}) \label{Hxbos}\\
&H_{mod} =4\bar \mu \cos (\theta_1)\cos (m\phi_{\bar 1}-m\phi).
\end{align}
Again, we will focus on the interface between the CDW dominated region $\Delta_{x}\neq 0$ for $x>0$ and the magnetic field dominated region  $\bar \mu\neq 0$ for $x<0$. Non-quadratic terms relevant in the renormalization group sense \cite{book_Luttinger}  lead to the chiral field $\theta_1$ being gapped uniformly throughout the system, say, $\theta_1=\pi \hat M$. Pinning of other fields is chosen in such a way that the total energy is minimized, so
\begin{align}
&\theta_{\bar 1} = \frac{\pi}{m}(\hat M +1+2\hat n),\ x>0,  \label{theta} \\
& \phi_{\bar 1} = \phi+\frac{\pi}{m}(\hat M +1+2\hat l) ,\  x<0, \label{phi}
\end{align}
where $\hat M$, $\hat n$, and $\hat l$ are integer-valued operators.
The only non-trivial commutation relation between them is $[\hat n, \hat l]=im/4\pi$. The corresponding zero-energy parafermion operator 
 \cite{PF_Clarke,PF_Mong,vaezi_2,PFs_Loss,PFs_Loss_2,PFs_TI} is given by
\begin{align}
&\alpha = e^{i \frac{2\pi}{m}  (\hat n+\hat l) }  , \ \alpha^{m}=1. \label{alpha}
\end{align}
The so found ground state is $m$-fold degenerate and the bound states are $\mathbb{Z}_m$-parafermions  obeying non-Abelian braiding statistics \cite{PF_Clarke,PF_Mong,PF_Cheng,PF_Linder,vaezi_2,PFs_Loss,PFs_Loss_2,PFs_TI,barkeshli_2}. This $m$-fold degeneracy can be explained following Refs. \cite{PF_Clarke,PF_Mong,PF_Cheng,PF_Linder}. Let us assume that we have a second interface at $x=L$ such that $\Delta_{x}= 0$ and $\bar \mu \neq 0$  for $x>L$.  The spin located in the gapped CDW dominated region ($0<x<L$) is given by $\left < s_z \right> =2(l_{<}-l_{>})/m$, assuming $m$ distinct values, where  $l_{\gtrless}$  is the  quantum number $\hat l$ right/left to this region. By analogy, the  magnetic field dominated region is characterized by the difference in the charge density between the two edges. The  parafermion relation is written as $\alpha_< \alpha_> =\alpha_> \alpha_< e^{2\pi i/m}$.

Similarly, the interface between the tunneling ($H_{tun}$) and field ($H_x$) dominated regions can also host fractional charges considered above. We note that $H_{tun}$ rewritten in terms of chiral fields becomes of the same form as Eq.~(\ref{Hxbos}) but with $\phi=0$.

{\it Exotic superconductivity.} Next we consider a TIC in proximity to a  bulk $s$-wave superconductor with phase ${ \phi_{sc}}$ 
and work in the spin filtering regime, see Fig.~\ref{spectrum}. 
The only proximity-induced superconducting term that can open a gap in the spectrum is of the form
\begin{align}
H_{sc}' = e^{i \phi_{sc}} \Delta_{sc}' R_{1}^\dagger L_{1}^\dagger R_{\bar 1}^\dagger L_{\bar 1}^\dagger + H.c.,
\end{align}
where $ \Delta_{sc}' $  is the proximity gap in the TIC. 
This term describes the coherent tunneling of {\it two} Cooper pairs out of the condensate of the bulk superconductor into the TIC edge states. The bosonized version of this term reads $H_{sc}' = 2\Delta_{sc}' \cos (2 \phi_1+ \phi_{sc})$ and
has  a minimum at $\phi_1=\pi(n+1/2) -\phi_{sc}$ for $ \Delta_{sc}' >0$  and $\phi_1=\pi n-\phi_{sc}$ for $ \Delta_{sc}' <0$, where $n$ is an integer. If it is a relevant term in the renormalization group sense~\cite{book_Luttinger}, it opens a partial gap in the spectrum, {\it i.e.}, the charge degrees of freedom are gapped out but the spin degrees of freedom stay gapless. The interface between two such regions forms a Josephson junction with a Josephson current of $8\pi m$-periodicity as a consequence of the $4m$-fold degeneracy of the ground state.

{\it Opening gaps by interactions.} Finally, we comment on the possibility of opening gaps in the spectrum solely via interactions. For example, the back-scattering exchange term $H_{int} = \Delta_{int}(R_1^\dagger L_{\bar 1} )(R_{\bar 1}^\dagger  L_1)+ H.c.$, which reads in
 bosonized form, 
\begin{align}
H_{int} = 2\Delta_{int} \cos (2\theta_1),
\end{align}
opens a gap in the charge sector. This exchange term requires overlap between the TIC edge states.
The interface between the region with $\Delta_{int}>0$ [$\theta_1=\pi (n+1/2)$] and the one with $\Delta_{int}<0$ [$\theta_1=\pi n$]  hosts a quasiparticle with fractional charge  $e/2m$.
The domain wall between $H_{sc}' $ and $H_{int}$ hosts  bound states in the charge sector. By analogy with the degenerate states considered above [see Eqs. (\ref{theta}) - (\ref{alpha})], the ground state in the charge sector is $2m$-fold degenerate.

The time-reversal invariant two-particle back-scattering term described in Refs. \cite{kane} and \cite{thomas} can result in the opening of a gap of the TI edges  and is given by $H_{um}=\int dx\ e^{-4ik_Fx} \psi_1^\dagger ( \partial \psi_1^\dagger)( \partial \psi_{\bar 1})\psi_{\bar 1} + H.c.$
In the bosonized form the term becomes
\begin{align}
&H_{um}=4 \Delta_{um} \cos (2 \theta_1) \cos(2 m \theta_{\bar 1}).
\end{align}
In contrast to $H_{int}$, this term opens a full gap in the spectrum with the  quasiparticle charge in the system given by $e/2m$  with $\theta_1 = \pi \hat M/2$ and $\theta_{\bar 1}= \pi ( \hat M+1+2 \hat n)/2m$. 
If the eigenvalue $M$ is an even (odd) number, $\left <s_y \right >\neq 0$ ($\left <s_y \right >= 0$) with $\left <s_x \right > = 0$ for all values of $M$. 
The interface between $H_{um}$ and $H_{mod}$ also hosts bound states. The charge field is pinned uniformly with $\theta_1=\pi \hat M$, $\phi_{\bar 1}$ is pinned according to Eq. (\ref{phi}) and $\theta_{\bar 1} =\pi (\hat n+1/2)/m$ replaces Eq. (\ref{theta}).
Again, two spin fields on different sides of the domain wall do not commute with each other, resulting in a degenerate ground state with non-Abelian statistics described by parafermion operators.

{\it Conclusions.} We have considered constrictions in two-dimensional topological insulators.
First, we show that such TICs could be used as spin filters operated solely by electric fields, {\it i.e.} by tuning the chemical potential. Second, we demonstrated that the proposed setup could be used to generate degenerate fractional bound states with non-Abelian statistics. The domain walls occur at the interfaces between regions of different gap-opening mechanisms. For example,  gaps can be opened by magnetic fields, tunneling between edges, charge density waves, or solely by interactions.  We finally note that the proposed coupled edge states  could be realized not only in TIs but also in systems of coupled wires \cite{yaroslav} in the framework of strip of stripes \cite{Lebed,Kane_PRL,Stripes_PRL,Kane_PRB,Stripes_arxiv,Stripes_nuclear,Neupert,
tobias_1,Oreg,yaroslav,gefen},  which is especially relevant for the fractional TI regime.

This work is supported by the Swiss NSF and NCCR QSIT.


\end{document}